\documentclass{ieeeaccess}
\usepackage{cite}
\usepackage{amsmath,amssymb,amsfonts}
\usepackage{algorithmic}
\usepackage{graphicx}
\usepackage{textcomp}

\usepackage{xcolor}

\usepackage{booktabs}
\usepackage{subcaption}

\def\BibTeX{{\rm B\kern-.05em{\sc i\kern-.025em b}\kern-.08em
    T\kern-.1667em\lower.7ex\hbox{E}\kern-.125emX}}
\begin{document}
\history{Date of publication xxxx 00, 0000, date of current version xxxx 00, 0000.}
\doi{10.1109/ACCESS.2017.DOI}

\title{Large-scale Mobile App Identification \\
Using Deep Learning}
\author{\uppercase{Shahbaz Rezaei}\authorrefmark{1}, \IEEEmembership{Member, IEEE},
\uppercase{Bryce Kroencke\authorrefmark{2}, and Xin Liu}.\authorrefmark{3},
\IEEEmembership{Fellow, IEEE}}
\address[1]{Computer Science Department, University of California, Davis, CA, USA (e-mail: srezaei@ucdavis.edu)}
\address[2]{Computer Science Department, University of California, Davis, CA, USA (e-mail: bakroencke@ucdavis.edu)}
\address[3]{Computer Science Department, University of California, Davis, CA, USA (e-mail: xinliu@ucdavis.edu)}

\markboth
{Author \headeretal: Preparation of Papers for IEEE TRANSACTIONS and JOURNALS}
{Author \headeretal: Preparation of Papers for IEEE TRANSACTIONS and JOURNALS}

\corresp{Corresponding author: Shahbaz Rezaei (e-mail: srezaei@ucdavis.edu).}

\begin{abstract}
Many network services and tools (e.g. network monitors, malware-detection systems, routing and billing policy enforcement modules in ISPs) depend on identifying the type of traffic that passes through the network. With the widespread use of mobile devices, the vast diversity of mobile apps, and the massive adoption of encryption protocols (such as TLS), large-scale encrypted traffic classification becomes increasingly difficult. In this paper, we propose a deep learning model for mobile app identification that works even with encrypted traffic. The proposed model only needs the payload of the first few packets for classification, and, hence, it is suitable even for applications that rely on early prediction, such as routing and QoS provisioning. The deep model achieves between $84\%$ to $98\%$ accuracy for the identification of 80 popular apps. We also perform occlusion analysis to bring insight into what data is leaked from SSL/TLS protocol that allows accurate app identification. Moreover, our traffic analysis shows that many apps generate not only app-specific traffic, but also numerous ambiguous flows. Ambiguous flows are flows generated by common functionality modules, such as advertisement and traffic analytics. Because such flows are common among many different apps, identifying the source app that generates ambiguous flows is challenging.  To address this challenge, we propose a CNN+LSTM model that uses adjacent flows to learn the order and pattern of multiple flows, to better identify the app that generates them. We show that such flow association considerably improves the accuracy, particularly for ambiguous flows. Furthermore, we show that our approach is robust to mixed traffic scenarios where some unrelated flows may appear in adjacent flows. To the best of our knowledge, this is the first work that identifies the source app for ambiguous flows.
\end{abstract}


\titlepgskip=-15pt

\maketitle

\section{Introduction}
\label{sec:introduction}
\PARstart{N}{etwork} traffic classification assigns labels to network traffic flows. Depending on its purpose, the labels can be various apps (e.g. Facebook, YouTube, and Skype), different traffic types (e.g. streaming, downloading, and browsing), user actions (e.g. liking or sharing), etc. Network traffic classification has been widely used for applying routing or billing policy, enforcing QoS requirements, identifying security threats or anomalies in firewalls or malware detection systems, etc. Because of its wide applications, traffic classification has been studied for more than two decades.

The earliest attempts to classify network traffic relied mostly on port numbers in TCP/UDP packets. Such approaches, however, are no longer sufficient because modern apps and protocols are often not tightly bound to a specific unique port number. The next set of approaches has focused on manually identifying patterns and keywords in unencrypted parts of the payload, called deep packet inspection (DPI). Nevertheless, such approaches are time-consuming and labor-intensive since they require continual manual extraction of signatures \cite{tongaonkar2015towards}.  Furthermore, their applicability has been reduced as more traffic is encrypted. 

To address the limitations of such methods, classical machine learning approaches have emerged. Classical machine learning approaches, such as support vector machine (SVM), K-nearest neighbor (KNN), etc., have achieved the state-of-the-art accuracy for many years. However, these approaches mostly rely on statistical features obtained from the entire flow which are not suitable for early classification, where the result of the classification is needed after observing the beginning of a flow. Furthermore, the rapid spread of mobile devices has dramatically changed the traffic pattern of the Internet. Mobile app identification poses a new challenge-the existence of a large number of applications. Due to the inability to capture more complex patterns for a large number of classes, the accuracy of traditional approaches has reduced. 

In recent years, deep learning has been widely adopted in various application scenarios from image recognition to text translation. A few recent studies show that deep learning models can achieve high accuracy for network traffic classification \cite{rezaei2018achieve,lotfollahi2017deep,rezaei2019multitask} and they do not require manual feature selection. However, these studies are limited to a small number of services or traffic types and their accuracy has yet to be investigated on large-scale mobile app identification.

In this paper, we focus on two main challenges in mobile app identification. First, we aim to solve the app identification problem for a large number of classes, generally known as extreme classification in machine learning. Although a few studies have already achieved acceptable accuracy in extreme classification, we perform an occlusion analysis to better understand the reason why deep models work. 
Second, we aim to identify source apps for \textit{ambiguous} traffic flows. Our analysis of the traffic shows that many apps generate not only app-specific flows, but also numerous ambiguous flows. Ambiguous flows are generated by embedded modules that are common among many different apps, such as advertisement, or common Web-API traffic flows \cite{taylor2017robust}. Identifying ambiguous traffic (i.e. labeling them as \textit{ambiguous}) is shown to be possible \cite{taylor2017robust}. However, identifying the source app that generates ambiguous traffic flows has not been studied before. The problem is challenging because ambiguous traffic can be generated by multiple apps and patterns in such traffic might be similar even when it is generated by different apps.

To tackle these challenges, we propose two deep learning models. We use a convolutional neural network (CNN) model for the identification of large scale mobile apps. Our dataset is comprised of 80 apps from a wide range of categories, including streaming, messaging, news, navigation, etc. We show that the proposed deep learning model achieves high accuracy across many network and encryption protocols for app identification by using payload information of the first few packets. 

To solve the second challenge of identifying the source app of an ambiguous flow, we develop a joint CNN and LSTM architecture building upon the developed CNN model. The model takes a few consecutive flows as input to identify the app that generates the first flow. The insight is that although one ambiguous traffic flow may not present any unique pattern, a set of consecutive flows that are generated by a single app presents unique patterns. The CNN part of the model captures the patterns in the header and payload of packets in each flow, while the LSTM part captures sequential patterns in adjacent flows. We show that our model is robust even if some flows belonging to other apps are taken into the model along with the app that we are interested in identifying. 

In summary, the contribution of this paper is as follows:

\begin{itemize}
	\item We show that a CNN model is capable of identifying mobile apps for 80-class classification with high accuracy.
	We perform fine-grained occlusion analysis for the first time on our proposed CNN model for traffic classification task. Our results bring insight into why a deep learning model can classify SSL and transport layer security (TLS) traffic flows. We show that certain handshake fields of SSL/TLS protocol leak enough information for app identification.
	\item Building upon the first CNN model, we propose a joint CNN+LSTM model to identifying the source app of ambiguous traffic flows as well as regular app-related traffic flows. Unlike previous studies where ambiguous traffic flows were removed or only detected as ambiguous, our joint model takes adjacent flows to identify the source app of the first flow and achieve high accuracy. To the best of our knowledge, this is the first work that takes the relation of several flows into account for identifying the source apps of ambiguous flows.
\end{itemize}

The remainder of this paper is organized as follows. Section \ref{sec-2} covers the most related work in literature about general traffic classification and smartphone app identification approaches. In Section \ref{sec-3}, we explain the detailed system architecture, including some definitions, input features, and deep model architectures. Section \ref{sec-4} contains our dataset description and evaluation of our models on real data. In Section \ref{sec-5}, we conclude the paper.

\section{Related Work}
\label{sec-2}
Although general traffic classification seems to be similar to mobile app identification, there are some differences in the types of traffic and patterns generated by smartphones that make them distinct from traditional traffic classification or website fingerprinting \cite{taylor2017robust, aceto2019mobile}. As a result, the studies in literature are divided into two groups: general traffic classification and mobile app identification.

\subsection{General Traffic Classification}

Classical machine learning approaches have been widely used in the past for network traffic classification \cite{velan2015survey}. Both supervised methods, including C4.5 \cite{alshammari2011can,alshammari2010investigation}, K-nearest neighbor \cite{bar2010realtime,wright2006inferring}, SVM \cite{kumano2014towards,khakpour2013information,kong2017identification}, naive Bayes \cite{okada2011application,sun2010novel} and unsupervised methods, such as Gaussian mixture model \cite{bernaille2007early} and K-means \cite{zhang2012encrypted,du2013design} have been used in the past. These methods often rely on statistical features of the entire flow, such as average packet length, standard deviation of inter-arrival time of packet, etc., for classification which is not suitable for \textit{early prediction}. Early prediction refers to scenarios where prediction is required as soon as traffic is observed by using the beginning of the flow. Routing, QoS provisioning, and malware-detection are just a few applications of traffic classification where early prediction is necessary. Moreover, the accuracy of such classical approaches have declined because of their simplicity, manual feature selection, and incapability of learning complex patterns \cite{rezaei2019deep}.

Recently, deep learning models have shown tremendous success in a wide range of applications from image recognition to text translation and speech recognition. The network community has also started applying deep learning models to tackle the network traffic classification task \cite{wang2019survey}. Statistical features of flows were converted into 2-dimensional arrays and used as the input of a CNN model in \cite{zhou2017method}. They successfully trained a model for a small number of classes for traffic type classification. However, it cannot be used for early prediction as the input is built based on the entire flow. In \cite{tong2018novel}, the authors used a combination of classical machine learning and deep learning for classification of five Google services that use QUIC protocol. They proposed a two step classification procedure: first, they used statistical features with random forest to classify three classes (that is, voice, chat, others), and then they used payload data with a CNN model for classifying other classes. This method is also not suitable for early prediction. A combination of CNN and stacked auto-encoders (SAEs) were used in \cite{lotfollahi2017deep} to classify traffic types and applications of ISCX dataset \cite{draper2016characterization}. Their approach achieved over $90\%$ accuracy for the classification of 17 apps. However, these studies focused on a handful of classes and treated deep learning models as black-box without investigating the features their models rely on.

In \cite{lopez2017network}, the authors investigated several combinations of deep models, including CNNs and long short-term memory (LSTMs), for classification of 15 services, including Office365 and Google. They also studied the combination of header and time-series features as input and their best model achieved about $96\%$ accuracy. In \cite{chen2017seq2img}, a CNN model was used to classify five protocols and five applications, separately. They converted time-series features into a 2-dimensional image by using reproducing kernel Hilbert space (RKHS). They achieved $99\%$ accuracy for 5-class classification. Interestingly, in their paper, SVM and decision tree also achieved over $97\%$ accuracy.

In \cite{rezaei2019deep}, a general guideline for applying deep learning models for the traffic classification was provided along with a general framework for capturing data, selecting a suitable deep model, and extracting proper input features. Although most traffic classification studies fall under the general framework, there is no concrete study to show that the framework works for large-scale classification problems with numerous classes.

In \cite{rezaei2018achieve}, the authors used a transfer learning approach to obviate the need for large labeled datasets. First, they trained a CNN model that takes the sampled time-series features of a flow and predicts several statistical properties of the flow, such as average packet length and average inter-arrival time. Then the model weights were transferred to a new model and the new model was trained on a small labeled dataset. The paper also showed that classification based on sampled time-series is feasible. In \cite{rezaei2019multitask}, the authors used a CNN model in multi-task learning framework to obviate the need for a large training dataset. They used a combination of a large unlabeled dataset and a small labeled dataset to train a model that predicts 3 tasks: application, bandwidth, and duration of flows. Both approaches were only studied on small-scale problems, with as few as 5 classes. Moreover, transfer learning is shown to have inherent security vulnerabilities \cite{rezaei2019target}.

Instead of classifying various traffic types with different underlying protocols, a few studies focused only on traffic of SSL/TLS protocols. In \cite{shbair2016multi}, the authors defined a new set of statistical features and used C4.5 and random forest to identify 9 services. In \cite{korczynski2014markov, shen2016certificate}, the authors used header information in SSL/TLS packets to create a fingerprint corresponding to a first-order \cite{korczynski2014markov} or second-order \cite{shen2016certificate} homogeneous Markov chain. They achieved high accuracy on a dataset of 12 web services. In \cite{pan2017wenc}, the authors extended the previous work by adding a weighted ensemble classifier to improve the accuracy. The model was also extended in \cite{liu2018mampf} to include length block sequence that considers the context of packets in a time-series manner to further improve the accuracy of fingerprinting. However, these approaches only work for SSL/TLS traffic and cannot be used for general mobile app identification where a wide range of protocols, including, SSL/TLS, unencrypted HTTP, RTP, SRTP, MTProto, etc., co-exist. Furthermore, these studies only considered a few classes and it is not shown to be scalable for a large number of classes.

\subsection{Smartphone Traffic Identification}

A wide range of analysis has been done on smartphone traffic \cite{conti2018dark}, including identifying users by the pattern generated from their set of installed apps \cite{stober2013you}, producing fingerprints of apps from their unencrypted HTTP traffic \cite{dai2013networkprofiler}, inferring sociological information (e.g. religion, health condition, sexual preference) \cite{barbera2013signals}, discovering the geographical position of a smartphone \cite{husted2010mobile},
detecting information leakage \cite{enck2014taintdroid}, identifying user actions \cite{conti2015analyzing, conti2015can}, etc.

In \cite{wang2015know}, the authors used the size and timing of 802.11 frames from WiFi traffic to identify 13 iOS apps. They used random forest and achieved over $90\%$ accuracy. However, it is only suitable for data captured on WiFi. In \cite{taylor2016appscanner, taylor2017robust}, AppScanner was introduced, which is an approach to identify mobile apps using statistical features extracted from the vector of the sizes of packets. They captured the traffic of 110 most popular apps in Google Store and achieved high accuracy. However, they simply removed ambiguous traffic without identifying them automatically. They extend AppScanner in \cite{taylor2017robust} by adding a module to identify ambiguous traffic and removing them automatically. However, they were not able to identify the source app of the ambiguous traffic. 

In \cite{alan2016can}, the authors used the website fingerprinting method, proposed in \cite{herrmann2009website, liberatore2006inferring}, to identify a large number of apps using three ML methods, including Gaussian naive Bayes and Multinomial naive Bayes. They only used the packet size of the first 64 packets and achieved maximum accuracy of $87\%$. Their dataset only contains Android traffic and it only works with captured data from WiFi. Despite their high accuracy on the collected dataset, they concluded that each device and Android/iOS version produces different packet size fingerprints. Hence, to have a practical model that uses their method, one should collect data from all combinations of devices and Android/iOS versions, which is impractical. Additionally, they completely ignored user activities and they only captured sessions with no meaningful interaction. In \cite{le2015antmonitor}, AntMonitor, a framework for collecting and analyzing network traffic, was introduced. They studied 70 Android apps. They used SVM and took 84 statistical features from traffic flows as input. They reported on F1 measure of $70.1\%$. However, both these methods are not suitable for early prediction. Furthermore, their dataset only contained very limited traffic types obtained only from a few Android devices.

Recently, deep learning has also been applied for mobile app identification. In \cite{li2017traffic}, a two-stage learning was proposed that performs unsupervised feature extraction in the first stage. The first stage does not need labeled data and can use public unlabeled data to improve accuracy. The second stage is a supervised category mapping using labeled data. They used variational auto-encoders and reached high accuracy for 12 apps.
In \cite{aceto2018mobile}, and their extended work \cite{aceto2019mobile}, the authors investigated several deep learning architectures, including 1-D CNN, 2-D CNN, LSTM, SAE, and multi-layer perceptron (MLP), to identify up to 49 mobile apps. They used the first \textit{N} bytes of the payload or both payload and header, depending on the setting, and reached up to $85\%$ accuracy for top-1 classification. They used the dataset containing 49 mobile apps generated by real user activities in \cite{aceto2018multi}.
In \cite{aceto2019mimetic}, Aceto et al. extended their previous work by introducing a novel framework that takes two different input viewpoints into account, namely data payload and protocol/time-series features. The multi-modal deep learning model can improve the performance by capturing patterns in both viewpoints.
In comparison with \cite{aceto2019mobile, aceto2019mimetic}, we show the applicability of deep learning models on even larger number of classes. Moreover, we perform occlusion analysis and identify ambiguous traffic flows, which have not been studied in previous work.

\section{System Architecture}
\label{sec-3}
In this section, we first introduce and define traffic objects. Then, we explain how extracted data from traffic objects is used as input features of our deep models. Then, we introduce our CNN and CNN+LSTM models which aim to solve mobile app identification and tackle ambiguous flows.

\subsection{Traffic Object}
Traffic object determines how raw captured packets are divided into multiple units with appropriate labels. 
The most common traffic object is a \textit{flow}. A flow is a set of packets that shares the same 5-tuple (i.e. source IP, source port, destination IP, destination port, and transport-layer protocol). In such a definition, direction is taken into account. For example, a TCP connection is divided into two flows: one containing packets in a forward direction (from client to server) and the other in a backward direction (from server to client). However, in most studies, the direction of a flow is ignored and all packets in both directions are considered as a part of a single flow, sometimes called \textit{biflow}. In this paper, we consider a flow with both directions because they achieve higher accuracy and also they essentially share the same label. For TCP connections, the beginning and the end of a connection are easily determined by TCP handshake packets, such as SYN and FIN. For other traffic types, such as UDP, a threshold (often between 15 to 60 seconds) is often used to indicate how long a flow can stay inactive before it is considered terminated.


By observing traffic flows of a single app, we notice that apps often generate flows that are common among others and they are not specific to the apps that generate them. This phenomenon is reported in \cite{taylor2017robust} as \textit{ambiguous} traffic. Ambiguous traffic includes advertisement traffic, third-party library traffic, etc. How to label such ambiguous traffic depends on the purpose of the classification. For instance, for billing purposes, it is preferred to identify the source application that generates the ambiguous traffic. 
In this paper, we conduct experiments with ambiguous traffic as well as without it to illustrate the effect of such traffic. More importantly, we also introduce a CNN+LSTM model that is specifically designed to improve identification of the source application of the ambiguous traffic.

\subsection{Input Features}
\label{sec-input-features}
Four types of input features are often used for traffic classification tasks \cite{rezaei2019deep}: 

\textbf{Statistical features:} These features represent statistical properties of the entire flow, such as average packet length or maximum inter-arrival time, and often require to observe the entire or a considerable portion of a flow. Note that in some cases such as \cite{kumano2014towards}, it was shown that statistical features can be computed based on the first \textit{K} packets and achieved the accuracy as high as using the entire flow, e.g. the first 50 packets for maximum packet size or the first 150 packets for average packet size. Although they showed that for some specific statistical features of a one-directional flow, the first 10 packets suffice, it is still larger than the first six packets used in this paper.

\textbf{Time-series features:} These features contain packet-related properties of a set of consecutive packets in a flow, including the packet length, inter-arrival time, and direction. Although most papers use the time-series features of the first few packets, consecutive packets from the middle of the flow and sampled packets have also been shown to achieve good accuracy \cite{rezaei2018achieve}.

\textbf{Header:} It refers to layer 3 and 4 header fields, such as port number, IP/TCP flags, protocol, etc. Nowadays, header fields alone often result in poor performance. For instance, the destination port number that was once heavily used for app classification leads to a poor performance, if used alone \cite{rezaei2019deep}. Hence, it is often used alongside payload or time-series features.

\textbf{Payload: } Payload data contains raw bytes of a packet that reside on layer 4 and above, containing application data, SSL handshake, etc. In the literature, the payload of the first few packets are often used for classification.

Statistical features are not suitable for online classification, or early classification, where the result of the classification should be available as soon as possible. In many applications, such as routing, QoS provisioning, or filtering, early classification is inevitable. The usefulness of time-series features heavily depends on the dataset and application within the dataset. We find that time-series features of the first few packets are not suitable for our classification goal. Note that the choice of input feature heavily depends on the dataset and one should try all possible ones to obtain the best model as suggested in \cite{rezaei2019deep}. In our early experiments, the accuracy of CNN, LSTM, and random forest models with time-series of the first 30 packets was considerably lower than our approach, as shown in Section \ref{sec-4}. Hence, time-series and statistical features are not good identifier for our dataset. We find that payload data with deep learning models leads to a significantly better model for our dataset, although they are computationally more resource-hungry than traditional machine learning algorithms.

Therefore, in this paper, we use header and payload information of the first 6 packets of a flow for classification. For each packet, we only use the first 256 bytes of header and payload. We find that using more bytes or packets barely changes our model's accuracy on our dataset.
If there are not enough data bytes in the packet, we pad the input data with zeros. We normalize each byte to constrain it within $[0,1]$. The first 256 bytes of the first six packets are concatenated and the input is represented by a 1 dimensional vector of length $1536$.

As far as input design is concerned, we have two options: 1) concatenate all 6 packets into a single channel, or 2) put each packet into a separate channel. In this paper, we chose the first approach for the following reason: For example, in the case of SSL/TLS, by analyzing the data, we find out that depending on how TCP handshake and SSL/TLS handshake goes, the SSL/TLS handshake packets can appear in different packets. For instance, SSL Hello often appeared in the third, forth, fifth, and sixth packets. Note that we do not remove duplicate Acks or re-transmitted packets during pre-processing because such pre-processing increases the computational complexity of an online classifier. Hence, if we use 6 different channels for each packet, the filters for each channel have to learn the same pattern all over again for each channel, which may need more data and computational resources. On the other hand, the only drawback of using a single channel and concatenating all packets is that some random patterns may appear at the position where two packets are concatenated. We empirically find the first option is slightly better, but the difference is negligible.

\subsection{Model Architecture for App Identification}
\label{sec-fisrt-cnn}

The goal of app identification is to assign a correct label to each flow only using the input features of that flow. In other words, the input of the deep learning model is the header and payload data of the first few packets of a flow, as explained in Section \ref{sec-input-features}, and the output is the corresponding class label. In our first model, we use convolutional neural networks (CNNs) as a deep learning model for app identification.

CNNs, inspired by visual mechanisms of living organs, work well with high dimensional inputs. Although multi-layer perceptrons (MLPs) are shown to be powerful, they fail to be efficiently trained on high dimensional input data which requires a large amount of hidden parameters to be trained. CNNs reduce the number of learning parameters significantly by using a set of convolution filters (kernels) in convolutional layers. These convolutional filters have only a few learning parameters and are shifted to cover the entire input. In other words, instead of learning parameters of each section of the input individually, the model learns a set of filters applied to the entire input. This also made a CNN model shift-invariant because the same filter can detect the same patterns regardless of its position on the input. 

\begin{figure}[h]
  \centering
  \includegraphics[width=\linewidth]{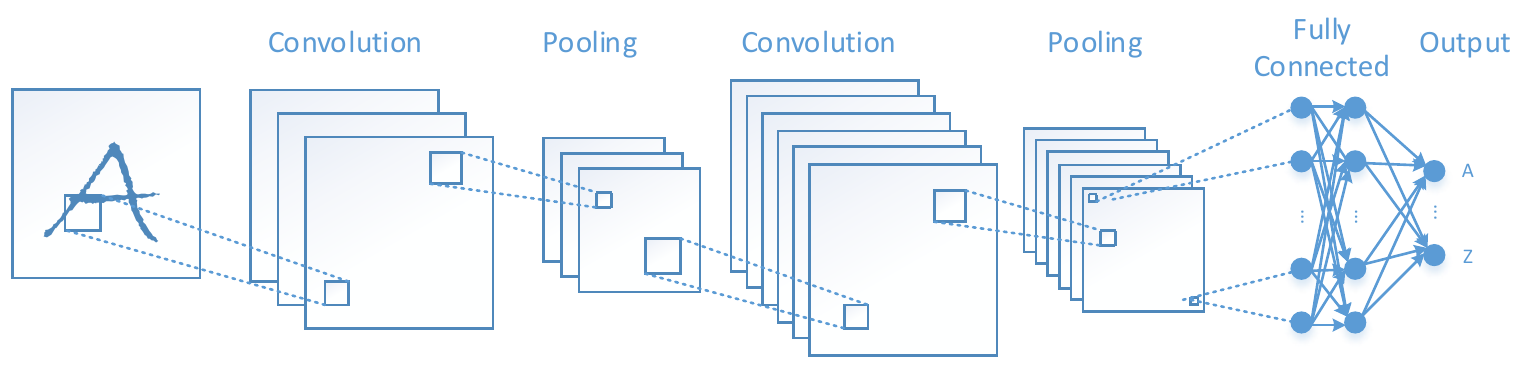}
  \caption{A typical CNN structure}
  \label{fig-cnn-sample}
\end{figure}

In addition to convolutional layers, CNNs also have pooling and fully connected (FC) layers. A typical CNN structure is shown in Figure \ref{fig-cnn-sample}. Pooling layers act as a non-linear down-sampling. The most common form of pooling is max pooling. In max pooling, the maximum values of each cluster of neurons are selected and used as an input of the next layer. Pooling layers reduce the size of their input and, consequently, reduce the number of training parameters and computation. FC layers are often used for the last few layers and they are known to perform the high level reasoning.

\begin{figure}[h]
  \centering
  \includegraphics[width=\linewidth]{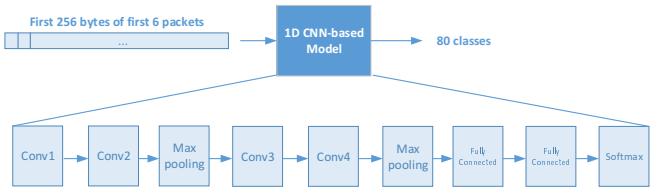}
  \caption{The proposed CNN model}
  \label{fig-multi-traffic-CNN}
\end{figure}

In this paper, we use a 1-dimensional CNN, as shown in Figure \ref{fig-multi-traffic-CNN}. We concatenate the 256 bytes of the first 6 packets into a single vector to feed to the CNN model. The reason we use a CNN model is that it is shift-invariant. This is crucial because the patterns that we find out to be useful for classification can appear almost anywhere in a packet. For example, TCP options can appear in a range of bytes, depending on the size of IP header and other TCP options. Moreover, app-specific data patterns can also appear anywhere in the payload. For example, we show that SSL/TLS handshake extensions are very useful in classification of SSL/TLS traffic and they can appear in a wide range of locations within a packet. In summary, because the location of such features are not predetermined in a packet, shift-invariant models are more suitable.

\subsection{Ambiguous Flow Identification}
\label{sec-model-lstm}
\begin{figure}[h]
  \centering
  \includegraphics[width=\linewidth]{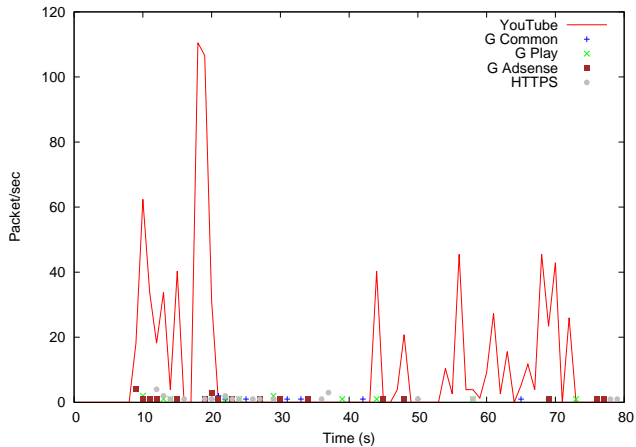}
  \caption{A sample of YouTube session with ambiguous flows}
  \label{fig-sample-multi-traffic}
\end{figure}

When a single app is running, such as YouTube, not all traffic flows are directly related to the content of the app. For example, some flows contain Google analytics or Ads data. Such traffic flows that are not specific to the  app and can also be generated by other apps are called ambiguous flows. A typical example of such a session is shown in Figure \ref{fig-sample-multi-traffic}. There are many other flows during the session. But, we only keep 5 flows to have a clear figure. These flows are generated as a result of the YouTube session. Hence, we are interested in identifying the source app that causes these flows to be generated. In other words, we want to label Google Ads and all other ambiguous flows as YouTube. However, the Google Ads traffic may also appear in other Google apps, such as Google Music, where we want to label (the source app of) Google Ads as Google Music when it generates the Google Ads traffic. Clearly, to identify the source app that generates the Google Ads traffic, only looking at the Google Ads traffic itself would lead to low accuracy.

To solve the problem, we propose a multiple-flow classification scheme where several subsequent flows are considered to identify the source app of the target flow. We call this problem \textit{flow association} because, unlike typical traffic classification, we exploit the relation of several adjacent flows all together. Here, the target flow refers to the flow that we are interested in identifying the source app and the source app refers to the app that generates the ambiguous target flow. Our intuition is that although a single ambiguous flow may not provide enough information to identify the source app, looking at the adjacent flows and observing their order, patterns, and relative starting time can help identify the source app. Hence, in our model, to identify the source app of a target flow, we also use the next $k$ flows. Note that in many cases, all adjacent $k$ flows that we feed to the model are ambiguous flows, but our insight is that each app generates a unique set of ambiguous flows with unique patterns and order. Furthermore, if the target flow is not an ambiguous flow, feeding the next adjacent flows should not hurt the accuracy.

Interestingly, Google services generate the most number of ambiguous traffic flows in our dataset. Hence, for the flow association task, we focus on 7 Google classes: Google Map, Google Music, Hangouts, Gmail, Google Earth, YouTube, and Google Play.
In addition to these traffic flows, Google apps also generate the following flows: Google Common, Google Analytics, Google Search, Google Adsense, TCP Connect, HTTP, and HTTPS. So, in total, we have 7 source apps and 7 ambiguous traffic labels. We are interested in identifying the source app of each flow. In other words, the final classification task is a 7-class classification (i.e. Google Map, Google Music, etc). 

To achieve this goal, we also use a type of recurrent neural networks (RNNs), called LSTM, which is suitable for sequential data. Such models have been successfully applied to speech recognition, translation task, time series prediction, and language modeling. The output of a LSTM model depends not only on the last input, but also on the previous inputs. In the flow association task, we can consider each flow as a single input and the set of consecutive flows as a sequence of input data which is suitable for LSTM.

\begin{figure}[h]
  \centering
  \includegraphics[width=\linewidth]{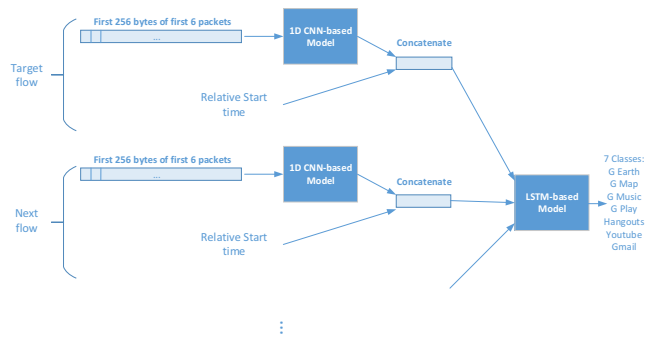}
  \caption{The joint CNN+LSTM model}
  \label{fig-multi-traffic-LSTM}
\end{figure}

To solve the problem of identifying source apps, we introduce a two-step training process. We first train a CNN model to predict the 14 true class labels using single flows. True class labels refer to the 7 Google apps and 7 ambiguous traffic labels altogether. We use the same CNN architecture and data input format as in Section \ref{sec-fisrt-cnn}, as shown in Figure \ref{fig-multi-traffic-CNN}. Next, we use an LSTM model to predict the source application of a flow. To do that, we feed not only the input features of the target flow, but also the input features of the next $k$ flows. The structure of the model is shown in Figure \ref{fig-multi-traffic-LSTM}. We first pass each flow input feature through the CNN model that predicts the true label. The weights of the CNN model are transferred from the first training step, shown in Figure \ref{fig-multi-traffic-CNN}. Then the predicted output is concatenated with the relative time of the flow (i.e. relative to the first/target flow). Then, it is fed as the first step of the LSTM model. We repeat the same process for the next $k$ flows as well. Then, the LSTM model is expected to predict the source app of the first flow.

We hypothesize that the flows related to a single session of an app start close to each other, as shown in Figure \ref{fig-sample-multi-traffic}. Hence, we define a threshold of 2 seconds to include adjacent flows. In other words, we only feed the CNN+LSTM model with the flows within the 2 seconds of the first flow. Otherwise, we pad the remaining flows input with zeros. The 2 seconds threshold also ensures that the traffic flows of two different apps are not mixed because it is highly unlikely that a user opens two different Google apps within two seconds. Hence, the flows that are fed to the CNN+LSTM model corresponds to only one app in most cases. However, in reality, the traffic flows of two different source apps may appear close to each other if one is run on the background. In Section \ref{sec-eval-lstm}, we evaluate the robustness of our approach to such mixed traffic scenarios.

\section{Evaluation}
\label{sec-4}
In this section, we describe our large mobile traffic dataset. Then, we evaluate our CNN-based model for app identification task with 80 classes. We use accuracy, precision, recall, and F1 to report the performance in different scenarios. We also illustrate the confusion matrix to show the most common misclassification patterns. Moreover, we perform occlusion analysis on the SSL/TLS encrypted traffic to shed lights on why apps using such an encrypted traffic protocol can still be identified. Finally, we evaluate our flow association approach using the CNN+LSTM model and illustrate it effectiveness in identifying the source app of ambiguous flows.

\subsection{Dataset Description}

The data was collected under controlled mobile operators networks in 2018\footnote{Due to the NDA with the data provider, we cannot reveal the name, network, and other details about the dataset.}. Two methodologies were used to obtain data:
1) Human users interacting with each app, where the data was captured by Wireshark. 2) Mobile device emulators where scripts emulated how users access the apps without human intervention. The human-generated portion of data and script-generated portion are roughly the same in terms of volume. The labels were assigned in a controlled environment where all unrelated apps were closed or removed.

\begin{figure*}[h]
  \centering
  \includegraphics[width=\linewidth]{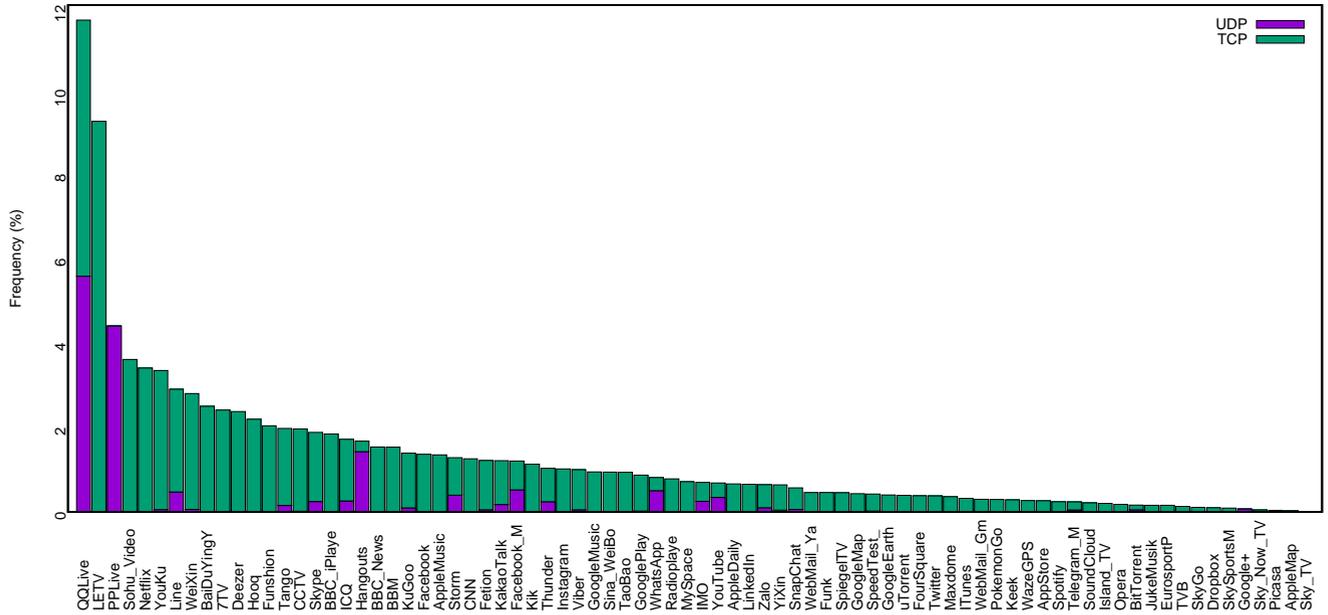}
  \caption{Frequency distribution of 80 classes.}
  \label{fig-frequency-dataset}
\end{figure*}

The dataset contains 80 mobile apps, including streaming, social media, messaging, Email, and navigation apps. It contains app traffic for both Android and iOS devices, except for apps that only have one version, such as Apple Map and Bit Torrent. Figure \ref{fig-frequency-dataset} shows the frequency of each application traffic in our dataset. As it is shown, the dataset is highly imbalanced. To mitigate the imbalance problem in the dataset, we use random under-sampling on the two most frequent apps (that is, QQLive and LETV) and random over-sampling on all apps with less than $1\%$ frequency. The reason we do not use more advanced over-sampling methods, such as Synthetic Minority Over-sampling Technique (SMOTE), is that our input feature is high dimensional. Since we feed the model with the raw packet payload without any feature selection, the data points of the same class maybe far away in the input space. So, methods such as SMOTE may create unrealistic synthetic points.

As shown in Figure \ref{fig-frequency-dataset}, the majority of the traffic flows are TCP and only around $10\%$ are UDP. Only streaming or video/voice applications contain a significant amount of UDP traffic, such as QQLive, PPLive, Facebook messenger, WhatsApp, Hangouts, etc. UDP traffic flows of Google services, such as YouTube and Hangouts, use Google QUIC protocol. Other UDP traffic mostly have RTP or SRTP (e.g. Skype and WhatsApp) or some custom protocols (e.g. MTProto for Telegram).

\begin{figure*}[h]
  \centering
  \includegraphics[width=\linewidth]{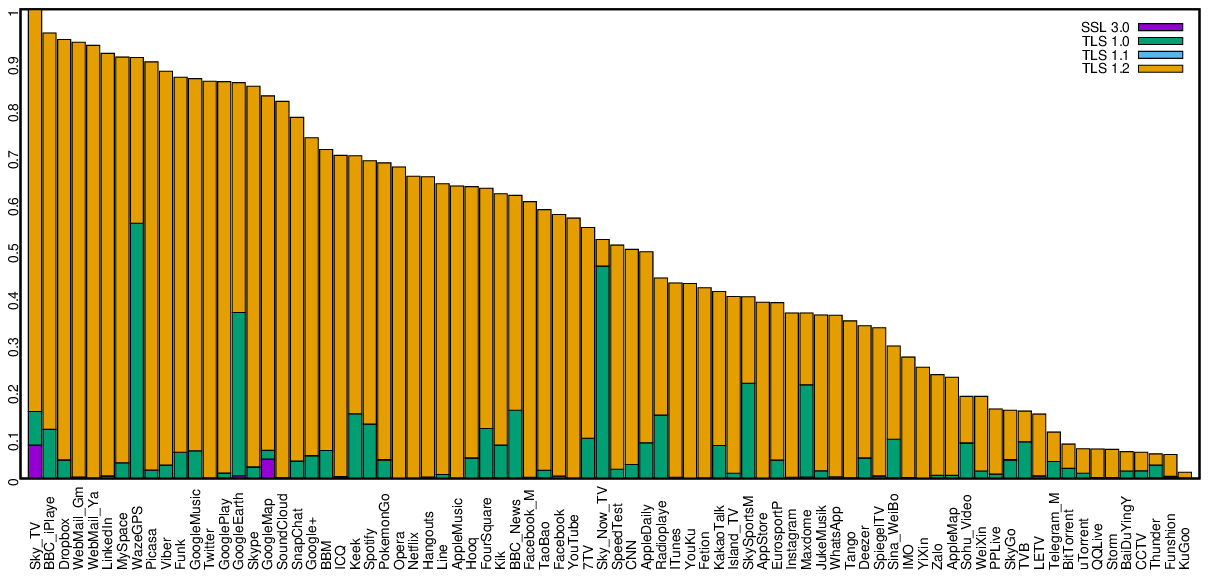}
  \caption{Portion of TCP flows encrypted with SSL/TLS}
  \label{fig-ssl-dataset}
\end{figure*}

TCP traffic flows contain plain HTTP, HTTPS and a small number of other protocols. HTTPS constitute $45.05\%$ of the TCP traffic flows and $40.36\%$ of the entire dataset. Figure \ref{fig-ssl-dataset} illustrates the amount of HTTPS traffic among TCP flows for each app. The traffic flows are shown based on encryption protocols and their version. Note that SSL and TLS are the primary encryption protocols used for HTTPS. While a SSL/TLS connection can be established on any port number, including port 80 \cite{razaghpanah2017studying}, we do not observe a significant amount of such flows. As it is shown, some applications (on the right side of the figure) heavily use HTTP instead of HTTPS. Moreover, some applications, such as Sky TV, still use SSL 3.0 which is deprecated. Furthermore, many applications, including Waze, Google Earth, Maxdome, still use TLS 1.0 traffic instead of the much stronger versions, such as TLS 1.2 or TLS 1.3. As it is reported in \cite{razaghpanah2017studying}, this is mainly a result of misconfiguration by developers because the dataset was captured using a wide range of OS versions that mostly support TLS 1.2.

\subsection{App Identification using CNN Model}
\label{sec-eval-single-flow}

\subsubsection{Classification without Ambiguous Traffic}

In this section, we train the CNN model, explained in Section \ref{sec-fisrt-cnn}, for the 80-class classification task. The details of the CNN model are summarized in Table \ref{tbl-model-specification}. Batch normalization is also used at the end of each pooling layer. We train the model with Adam optimization algorithm with maximum of 30 epochs and $0.001$ as the learning rate. We use early stopping, similar to \cite{lopez2017network}, where the training is terminated if the loss function does not improve for 5 epochs. Moreover, the whole training data is used during one epoch. Our results are based on 10-fold cross-validation. In the first experiment, we remove all ambiguous traffic from the dataset. It has been shown that it is possible to remove ambiguous traffic automatically in \cite{taylor2017robust}.

To compare our approach with the state-of-the-art methods, we also conduct experiments with deep learning models on time-series features, similar to \cite{lopez2017network}, and random forest on statistical features, based on \cite{aceto2018mobile}. In \cite{lopez2017network}, several models, including LSTM (called RNN in \cite{lopez2017network}), CNN and various combinations of them, are trained on time-series features. We use their CNN, RNN, and their best performed model, namely CNN+RNN-2, in our experiment. We use direction, payload size, inter-arrival time, and window size (only in TCP packets). Furthermore, we train a random forest (RF) model similar to the experiment in \cite{aceto2018mobile}. The RF model takes 40 carefully handcrafted features obtained from the entire flow \cite{taylor2016appscanner}. Other details of our experiments are similar to \cite{lopez2017network} and \cite{aceto2018mobile}.

\begin{table*}[t]
	\centering
	\caption{Structure of the CNN model}
	\label{tbl-model-specification}
	\begin{tabular}{*{15}{c}}
		\hline
		- & Conv1 & Conv2 & Pool3 & Conv4 & Conv5 & Pool6 & FC7 & FC8 \\
		\hline
		Number of filters/neurons & 256 & 256 & - & 128 & 128 & - & 128 & 128 \\
		\hline
		Kernel size & 3 & 3 & 2 & 2 & 2 & 2 & - & - \\
		\hline
		stride size & 1 & 1 & 2 & 1 & 1 & 2 & - & - \\
		\hline
	\end{tabular}
\end{table*}

The results of the CNN model are shown in Table \ref{tab-exp1-cnn}. The highest accuracy is $99.12\%$ when the entire 256 bytes in the first 6 packets are used. However, due to the limited number of IP addresses in the dataset, the model may overfit to non-related features. Hence, we removed the layer 2 and 3 data and further evaluated the model's accuracy. Moreover, we also removed layer 4 data as well to investigate the effect of port numbers and TCP options on accuracy. In these cases, we still used the first 256 bytes of the first six packets, but we remove the above mentioned layers, respectively.
As shown in Table \ref{tab-exp1-cnn}, by only using the payload data without any header data, we can obtain $94.22\%$ accuracy. This clearly illustrates that mobile apps leave enough fingerprints for identification. Moreover, the accuracy of identifying apps are $96.87\%$, $80.64\%$, for HTTP and HTTPS traffic, respectively. This shows that even encrypted traffic has enough information for a reasonable accuracy. As mentioned in Section \ref{sec-input-features}, payload data is more suitable for classification of our dataset. That is why RNN-1, CNN-1, and CNN+RNN-2 that use time-series features perform poorly. Statistical features are also not accurate in our dataset as random forest that uses them performs worst. We analyze how our model can classify encrypted traffic in Section \ref{sec-occlusion}.

\begin{table}
  \caption{80-class classification without ambiguous traffic}
  \label{tab-exp1-cnn}
  \begin{tabular}{ccccl}
    \toprule
    Input detail & Precision & Recall & F1 & Accuracy\\
    \midrule
    All layers & 99.53\% & 98.72\% & 99.03\% & 99.12\% \\
    L 2/3 removed  & 98.36\% & 95.58\% & 96.78\% & 96.95\%\\
    L 2/3/4 removed & 96.99\% & 93.22\% & 94.99\% & 94.22\%\\
    RNN-1 \cite{lopez2017network} & 81.96\% & 74.36\% & 77.97\% & 78.55\%\\
    CNN-1 \cite{lopez2017network} & 71.84\% & 70.04\% & 70.92\% & 71.27\%\\
    CNN+RNN-2 \cite{lopez2017network} & 82.43\% & 78.11\% & 80.21\% & 80.22\%\\
    RF \cite{aceto2018mobile} & 64.21\% & 59.05\% & 61.52\% & 63.58\%\\
  \bottomrule
\end{tabular}
\end{table}

\subsubsection{Classification with Ambiguous Traffic} In the second experiment, we keep the ambiguous traffic flows and the goal is to label them based on their source apps that generate them. This is a significantly harder task since certain types of ambiguous flows, such as advertisement traffic, are often generated by multiple apps and they might not have enough distinguishable features to properly identify the source app. Table \ref{tab-exp2-cnn} presents the performance metrics of the second experiment. By comparing the results to the first experiment, we find that the presence of ambiguous traffic degrades the accuracy considerably. However, the accuracy of the model without header data is $84.01\%$ which is still higher than previous studies for large-scale app identification, such as \cite{taylor2017robust} with the highest accuracy of $73.5\%$ for a 65-class classification and similar settings. The biggest difference between our study and \cite{taylor2017robust} is that we use payload data for classification whereas they use time-series features. This shows that, even in the presence of encryption protocols, payload data still contains information that can improve the classification accuracy.

\begin{table}
  \caption{80-class classification with ambiguous traffic}
  \label{tab-exp2-cnn}
  \begin{tabular}{ccccl}
    \toprule
    Input detail & Precision & Recall & F1 & Accuracy\\
    \midrule
    All layers & 96.39\% & 94.67\% & 95.52\% & 95.90\% \\
    L 2/3 removed  & 90.55\% & 90.12\% & 90.32\% & 90.35\%\\
    L 2/3/4 removed & 84.81\% & 83.21\% & 83.99\% & 84.01\%\\
    RNN-1 \cite{lopez2017network} & 70.32\% & 67.12\% & 68.68\% & 69.61\%\\
    CNN-1 \cite{lopez2017network} & 63.27\% & 59.94\% & 61.56\% & 62.22\%\\
    CNN+RNN-2 \cite{lopez2017network} & 74.82\% & 70.74\% & 72.72\% & 72.97\%\\
    RF \cite{aceto2018mobile} & 61.83\% & 59.29\% & 60.53\% & 61.24\%\\
  \bottomrule
\end{tabular}
\end{table}

\begin{figure}[h]
  \centering
  \includegraphics[width=\linewidth]{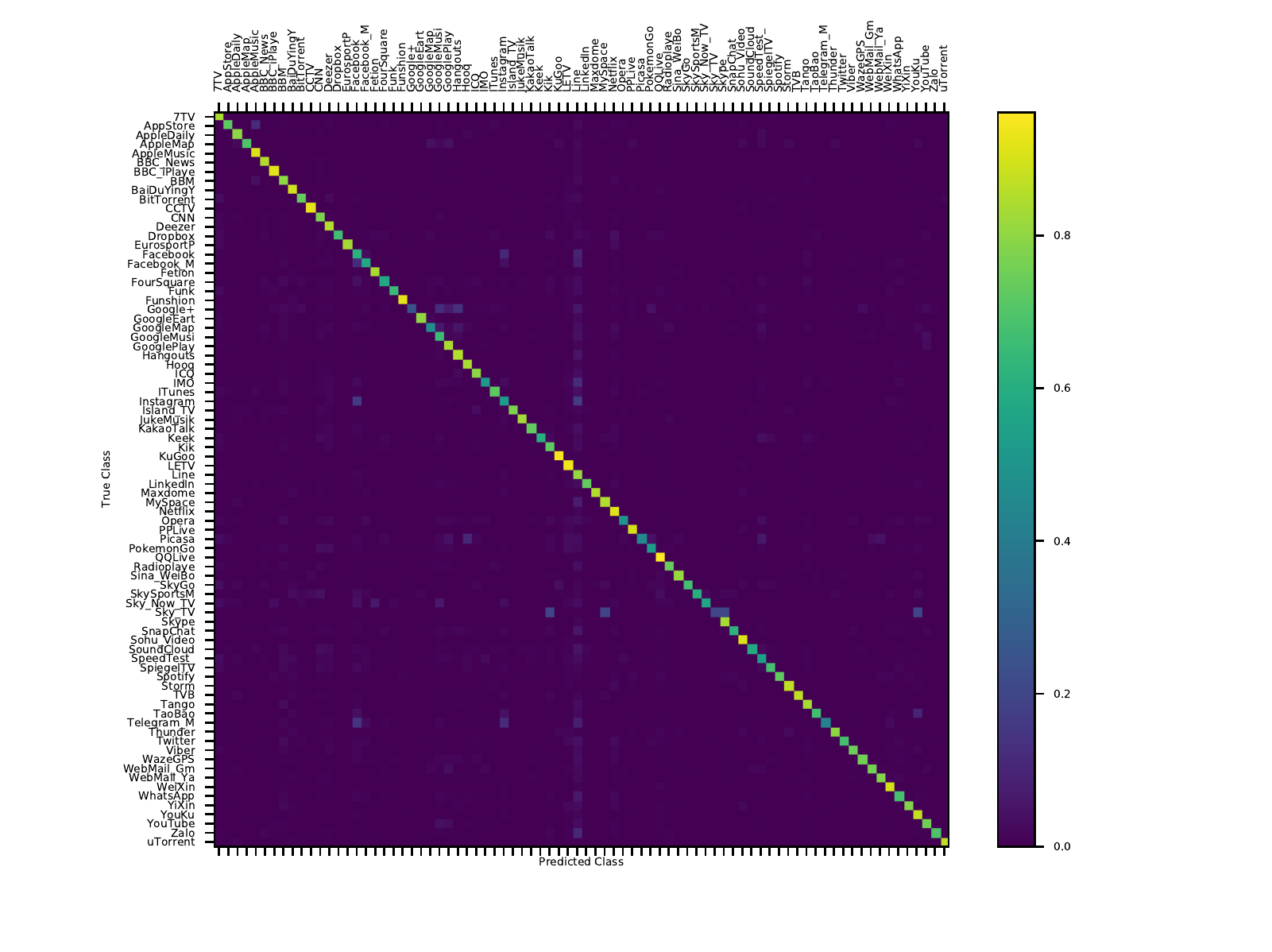}
  \caption{Confusion matrix of the 80-class classifier}
  \label{fig-cm-all-traffic}
\end{figure}

Accuracy across different classes varies. Figure \ref{fig-cm-all-traffic} shows the confusion matrix of the second experiment. Among all classes, Sky TV, Telegram, Picasa, and Google+ are the hardest to classify. These apps are amongst the classes with lowest number of samples. Moreover, according to Figure \ref{fig-ssl-dataset}, Sky TV and Picasa are amongst the classes with highest ratio of SSL and TLS traffic which make them more difficult to identify.

Sometimes the traffic patterns are different for different versions of an app, particularly between Android and iOS versions. To see whether classification accuracy is different for iOS and Android, we test the performance separately. Table \ref{tab-ios-android} presents the performance metrics of the two versions with a model trained only on the payload data. Although our dataset contains more samples of Android apps, the accuracy, recall, and precision of the two versions are not significantly different.

\begin{table}
  \caption{Performance of iOS vs Android (L 2/3/4 removed). Dist stands for distribution of the traffic within the entire dataset.}
  \label{tab-ios-android}
  \begin{tabular}{cccccl}
    \toprule
    Device & Dist & Precision & Recall & F1 & Accuracy\\
    \midrule
    iOS & 43.61\% & 84.11\% & 83.35\% & 83.74\% & 83.41\%\\
    Android & 56.39\% & 85.31\% & 83.98\% & 84.63\% & 84.54\%\\
  \bottomrule
\end{tabular}
\end{table}

We also test the performance on different types of traffic. As shown in Table \ref{tab-traffic-types}, the highest performance is obtained for unencrypted HTTP which is around $45\%$ of the entire dataset. UDP traffic covers about $10\%$ of the entire dataset and the accuracy is lower than the unencrypted HTTP. HTTPS traffic, covering around $40\%$ of the entire dataset, has the lowest accuracy. Although, the encrypted traffic should supposedly not reveal information about the content of the traffic, the accuracy is still significantly high ($75.43\%$) for 80-class classification. In Section \ref{sec-occlusion}, we analyze the features that allow the classification of the SSL/TLS traffic by conducting occlusion analysis.

\begin{table}
  \caption{Performance of different traffic types (L 2/3/4 removed)}
  \label{tab-traffic-types}
  \begin{tabular}{cccccl}
    \toprule
    Traffic & Dist & Precision & Recall & F1 & Accuracy\\
    \midrule
    TCP &  89.58\% & 84.46\% & 83.52\% & 83.99\% & 83.72\%\\
    UDP & 10.41\% & 88.25\% & 86.71\% & 87.47\% & 86.76\%\\
    HTTP & 45.58\% & 91.72\% & 91.43\% & 91.57\% & 91.60\%\\
    HTTPS & 40.36\% & 76.66\% & 74.53\% & 75.58\% & 75.43\% \\
  \bottomrule
\end{tabular}
\end{table}

\begin{figure}[h]
  \centering
  \includegraphics[width=\linewidth]{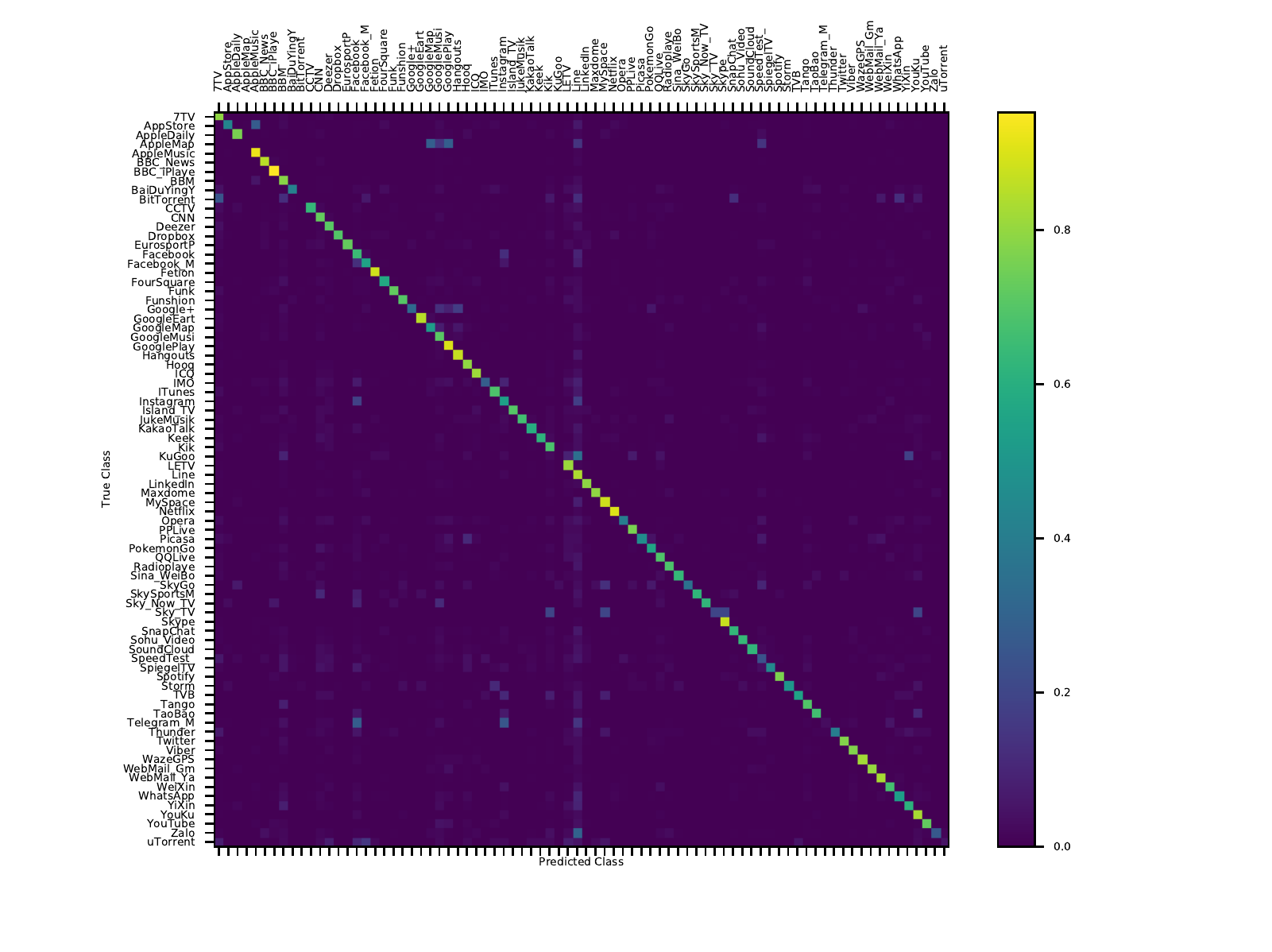}
  \caption{Confusion matrix of the 80-class classifier on HTTPS traffic}
  \label{fig-cm-ssl-tls}
\end{figure}

Figure \ref{fig-cm-ssl-tls} illustrates the confusion matrix of only the HTTPS traffic. The majority of the other apps are classified correctly by our model even for HTTPS traffic, as shown in Figure \ref{fig-cm-ssl-tls}. The recall of the AppleMap, and BitTorrent classes were almost zero because both of these classes have a small number of samples and among them only a small portion of samples are HTTPS. KuGoo's recall for HTTPS is also extremely low since it has the lowest number of HTTPS samples ($0.042\%$ of all HTTPS flows). Note that the overall recall of these classes are high because the majority of their flows are non-HTTPS and our model can correctly identify them. The recall of the AppStore, BaiDuYingY, Google+, IMO, Opera, Picasa, SkyGo, SkyTV, SpeedTest, Telegram, Thunder, Zalo, and uTorrent are also less than $60\%$. Although, some of these apps have many HTTPS samples, their low recall shows that they have similar features to other applications that make the classification difficult. For example, IMO traffic is often confused with Instagram and Facebook.

\subsection{Occlusion Analysis}
\label{sec-occlusion}

In this section, we perform occlusion analysis to reveal the features that allow the model to classify HTTPS encrypted traffic. We do not analyze the unencrypted HTTP traffic because deep packet inspection (DPI) methods have been shown to be effective on unencrypted traffic and it is trivial that payload information in traffic such as HTTP provides significant information about the apps. Moreover, for occlusion analysis of UDP traffic, we need to know the exact protocol (e.g. RTP, SRTP, DTLS, etc.,) used in each flow to be able to associate each byte of the payload to the protocol's fields. However, it is not possible for the majority of UDP traffic flows that are captured in the middle of the route to do such analysis. Hence, we only perform occlusion analysis for HTTPS traffic in this paper.

Occlusion analysis has been used extensively in image recognition task to show the sensitivity of CNN-based models to different components of an input image \cite{zeiler2014visualizing}. Typically, it is done by blocking a part of an image at inference time to see how it affects the classification accuracy. For example, one can block the trunk of an elephant on animal classification task, to see how much the model is sensitive to the trunk for classification.

In this section, we replace different parts of SSL/TLS handshake fields with a default value at inference time to see the effect of that fields on the classification accuracy. Note that not all traffic flows contain all SSL handshake fields. Moreover, each handshake field can appear at a wide range of bytes within a packet, depending on the size of previous layers and order and existence of SSL/TLS extensions. Hence, we wrote a parser that parses the entire packet and finds the target SSL/TLS fields and replaces it with a default value (zeros).

\begin{table}
  \caption{80-class classification occlusion analysis}
  \label{tab-occlusion-cnn}
  \begin{tabular}{cl}
    \toprule
    Occluded part & Accuracy\\
    \midrule
    - & 75.43\% \\
    Version & 75.34\% \\
    Record+Handshake Length & \textbf{59.31\%} \\
    Handshake type & 75.44\% \\
    Rand (Client Hello) & 71.40\% \\
    Rand (Server Hello) & 75.31\% \\
    SID info (Client Hello) & 71.84\% \\
    SID info (Server Hello) & 75.30\% \\
    Compression info (Client Hello) & 75.43\% \\
    Compression info (Server Hello) & 75.43\% \\
    Cipher info (Client Hello) & \textbf{51.67\%} \\
    Cipher info (Server Hello) & 75.36\% \\
    Certificate info (type 11) & 74.39\% \\
    Server key exchange info (type 12) & 75.43\% \\
    Client key exchange info (type 16) & 75.43\% \\
    All extensions & \textbf{27.52\%} \\
    Extension \#0 (SNI) & \textbf{38.19\%} \\
    Extension \#5 (Status request) & 75.40\% \\
    Extension \#10 (Supported Group/Elliptic Curves) & 72.20\% \\
    Extension \#11 (Elliptic curve point format) & 74.93\% \\
    Extension \#13 (Signature algorithm) & 75.01\% \\
    Extension \#16 (App layer protocol negotiation) & 72.26\% \\
    Extension \#18 (Signed certificate timestamp) & 75.12\% \\
    Extension \#23 (Extended master secret) & 74.87\% \\
    Extension \#35 (Session ticket) & 73.97\% \\
    Extension \#54-65279 (Unassigned) & \textbf{66.32\%} \\
    Extension \#65281 (Renegotiation info) & 72.36\% \\
    Cipher info (Client), All extensions & 12.42\% \\
    Cipher info (Client), All extensions, Rand (Client) & 8.20\% \\
  \bottomrule
\end{tabular}
\end{table}

Table \ref{tab-occlusion-cnn} presents the result of the occlusion analysis on the HTTPS traffic using SSL/TLS. Handshake fields with the most significant impact on the accuracy are highlighted. Extension number zero, or Server Name Indication (SNI), is the most important field in the SSL/TLS handshake. The importance of SNI is not surprising since it contains the destination hostname a client wants to connect and sometimes it is even used to obtain the ground truth \cite{shbair2016multi}. Cipher info is also useful for classification because it covers a wide range of cipher suite's combinations that may reveal target apps. Interestingly, SSL/TLS length also helps app identification. This is because different combinations of the extensions lead to different SSL/TLS length values and may reveal target apps. 

Interestingly, unassigned extensions (number 54 to 65279) also have an impact on the accuracy. It is reasonable because these extensions are used in a custom way by apps and it is highly unlikely that two apps use the same unassigned extension. So, for apps that use unassigned extensions, it is an important feature for classification. We observe unassigned extensions across 7 different mobile applications, including Gmail, Tango, and BBM.

It is interesting to see that removing the random number in the client hello also slightly degrades the accuracy. Note that the random field in the client hello packet consists of GMT time and a random number. In our dataset, some applications are captured at only a few time instances. As a result, we believe that the GMT part of the random number is probably exploited by the model. In a completely unbiased dataset, the client random number should not affect the performance of the model.

Our analysis is mostly based on client hello and server hello packets. Note that we only use the first 6 packets of each flow for classification. As a result, we mostly feed the model with the client and server hello packets. Only a small portion of flows has server certificate, server key exchange, or certificate request in their first 6 packets, but we found out that it did not have a significant role in the model's accuracy. Note that our model achieved acceptable accuracy for 80-class classification. Hence, we did not take more packets into input features to avoid making the model highly complex for a negligible performance gain. Moreover, using fewer packets for classification is more suitable for early (online) classification. That is the reason we only report data fields of client and server hello packets in Table \ref{tab-occlusion-cnn}.

\subsection{Source App Identification for Ambiguous Traffic}
\label{sec-eval-lstm}

\subsubsection{Multiple Flows Evaluation}
In this section, we show that our flow association approach with multiple flows can better solve the identification of source apps of ambiguous flows. In this section, we only focus on 7 Google services that generate the most ambiguous traffic flows, that is, Google Map, Google Music, Hangouts, Gmail, Google Earth, YouTube, and Google Play. Although we show that 80-class classification is achievable, one can easily divide traffic flows into several groups based on the IP addresses. For instance, Google traffic flows can be identified by the pool of Google IP addresses. However, the IP address is not accurate for fine-grained classification of apps. Hence, we still need a classifier that looks into payload or other features. In all the evaluations in this section, we remove all layer 2 to layer 4 data headers.


Similar to Section \ref{sec-eval-single-flow}, we first train the CNN model, described in Section \ref{sec-fisrt-cnn}, for Google service classification. 
The training procedure for the CNN part is explained in Section \ref{sec-fisrt-cnn}. We use a single LSTM layer with 50 units and dropout ratio of $0.5$. For the CNN+LSTM model, we re-train the entire model with Adam optimization algorithm with maximum of 50 epochs. Other training parameters are similar to Section \ref{sec-fisrt-cnn}.
The performance results are shown in Table \ref{tab-cnn-lstm}. As it is shown, the single-flow classification only achieves $91.98\%$. Note that RNN-1, CNN-1, and CNN+RNN-2 models take time-series features as input, whereas our CNN and CNN+LSTM models take payload data.

\begin{table}
  \caption{7-Google-class classification with ambiguous traffic (Thr stands for threshold)}
  \label{tab-cnn-lstm}
  \begin{tabular}{cccccl}
    \toprule
    Model & Thr & Precision & Recall & F1 & Accuracy\\
    \midrule
    CNN & - & 92.01\% & 91.97\% & 91.99\% & 91.98\%\\
    \midrule
    CNN+LSTM & 0.5 & 95.37\% & 95.01\% & 95.19\% & 95.08\%\\
    \midrule
    CNN+LSTM & 1 & 95.88\% & 95.91\% & 95.89\% & 95.91\%\\
    \midrule
    CNN+LSTM & 2 & 96.35\% & 96.23\% & 96.25\% & 96.23\%\\
    \midrule
    CNN+LSTM & 3 & 96.30\% & 96.23\% & 96.26\% & 96.29\%\\
    \midrule
    CNN+LSTM & 4 & 96.32\% & 96.19\% & 96.25\% & 96.29\%\\
    \midrule
    RNN-1 \cite{lopez2017network} & - & 77.17\% & 75.47\% & 76.31\% & 76.41\%\\
    \midrule
    CNN-1 \cite{lopez2017network} & - & 71.38\% & 68.06\% & 69.67\% & 69.97\%\\
    \midrule
    CNN+RNN-2 \cite{lopez2017network} & - & 82.56\% & 81.08\% & 81.81\% & 82.13\%\\
    \midrule
    RF \cite{aceto2018mobile} & - & 70.87\% & 66.24\% & 68.47\% & 68.89\%\\
  \bottomrule
\end{tabular}
\end{table}

To improve the accuracy of ambiguous flows, we conduct an experiment, explained in Section \ref{sec-model-lstm}. We first train the CNN model to classify all 14 classes, including both Google apps labels and ambiguous class labels, shown in Figure \ref{fig-multi-traffic-CNN}. Then, we transfer the weight of the CNN model to the CNN+LSTM model, shown in Figure \ref{fig-multi-traffic-LSTM}. Then, we train the CNN+LSTM model with multiple flows as described in Section \ref{sec-model-lstm}. For each flow, we find that the next two flows (within two seconds of the target flow) are enough for multiple-flow classification with high accuracy. So, we feed the CNN+LSTM model with only three flows. The performance of the CNN+LSTM model is shown in Table \ref{tab-cnn-lstm}. The multiple-flow classification improves the accuracy more than $4\%$ which is significant since the majority of Google traffic is encrypted. There is a trade-off between accuracy and how early we can make a prediction. As the time threshold increases, the accuracy improves. However, it is also desirable to classify apps as early as possible. Given the accuracy does not significantly improve for thresholds that are larger than 2 seconds, we choose the 2-second threshold. Note that there is a higher chance that flows are generated from different apps if we consider two far away flows in the time domain. Hence, we choose a relatively small threshold that does not affect accuracy too much.

\subsubsection{Robustness Analysis}
In reality, however, traffic flows of other Google apps may appear close to the target flow. In such cases, the second or third flows that are fed to the CNN+LSTM model may not have the same source apps as the first/target flow. To evaluate the robustness of our model to such a mixed traffic flows, we conduct the following experiment with noisy data: First, we duplicate each sample (containing 3 adjacent flows) in our dataset. Then, for each duplicate sample, we randomly replace a flow in our multiple-flow input with a random flow from another source app. The modified samples mimic traffic when captured in the real world where apps are not run in isolation. The results are shown in Table \ref{tab-cnn-lstm-robustness}. The number after \textit{R} represents the flow number that was replaced: R3 means only the third flow is replaced, R2 means only the second flow is replaced, and R23 means both second and third flows are replaced with random flows of another app. In Table \ref{tab-cnn-lstm-robustness}, \textit{T} represents the scenarios where we also train the model with replaced flows to mimic the flows in a real network. 

As is shown, the model is pretty robust when only one flow (out of three flows) is replaced with another app flows (R2 and R3 scenarios). However, when two flows are replaced (R23 scenario), the accuracy of the CNN+LSTM model with multiple flows becomes lower than the CNN model with a single flow (Table \ref{tab-cnn-lstm}). This is reasonable because it adds noisy data at inference time without training the model to handle such noisy data. However, if we train the model to deal with such a mixed traffic scenario, the model outperforms single flow classification in all scenarios of mixed flows (R2+T, R3+T, and R23+T). That is probably because the model learns to identify mixed traffic and it only uses the second and third flows when it is related to the first flow. Therefore, our flow association formulation with CNN+LSTM model always achieves highest accuracy for app identification with ambiguous traffic.

\begin{table}
  \caption{Robustness of the CNN+LSTM model to mixed traffic flows. The number after \textit{R} represents the flow(s) number that was replaced. \textit{T} represents the scenarios where the model trained with the replaced flows to mimic the flows in a real network.}
  \label{tab-cnn-lstm-robustness}
  \begin{tabular}{cccccl}
    \toprule
    Scenario & Direction & Precision & Recall & F1 & Accuracy\\
    \midrule
    R3 & Backward & 95.34\% & 94.85\% & 95.08\% & 95.09\%\\
    \midrule
    R2 & Backward & 95.11\% & 94.86\% & 94.98\% & 94.99\%\\
    \midrule
    R23 & Backward & 91.20\% & 91.06\% & 91.10\% & 91.10\%\\
    \midrule
    R3+T & Backward & 96.15\% & 95.98\% & 96.06\% & 96.03\%\\
    \midrule
    R2+T & Backward & 95.25\% & 95.76\% & 95.49\% & 95.74\%\\
    \midrule
    R23+T & Backward & 95.59\% & 95.41\% & 95.49\% & 95.50\%\\
    \midrule
    
    R3 & Forward & 95.05\% & 94.99\% & 94.01\% & 95.00\%\\
    \midrule
    R2 & Forward & 95.02\% & 94.82\% & 94.86\% & 94.89\%\\
    \midrule
    R23 & Forward & 90.60\% & 90.51\% & 90.55\% & 90.55\%\\
    \midrule
    R3+T & Forward & 96.03\% & 95.91\% & 95.96\% & 95.99\%\\
    \midrule
    R2+T & Forward & 95.46\% & 95.30\% & 95.36\% & 95.38\%\\
    \midrule
    R23+T & Forward & 95.11\% & 94.56\% & 94.83\% & 94.85\%\\
  \bottomrule
\end{tabular}
\end{table}

Note that during the evaluation, we find that if we feed the CNN+LSTM model with the sequence of flows in reverse order, we achieve better accuracy. In other words, we first take the third flow, then the second next flow, and then the first/target flow as input to the model. 
As it is shown in Table \ref{tab-cnn-lstm-robustness}, the accuracy of the model is slightly higher when it is fed with flows in backward direction, particularly when we train the model with noisy data.
The reason is related to how forget gates work in LSTM models. These gates help the model forget the previous hidden representation if they receive particular kinds of input. In many cases, particularly in the mixed traffic scenario, the most relevant flow to identify the source app is the first/target flow, although it might be ambiguous. So, by feeding the first/target flow last, we allow the LSTM model to forget the second and third flows if they are unrelated. All the results in this section are for the input in reverse order. However, to avoid confusion, we still use the first flow to refer to the target flow in this section.

\begin{figure}[h]
  \centering
  \includegraphics[width=\linewidth]{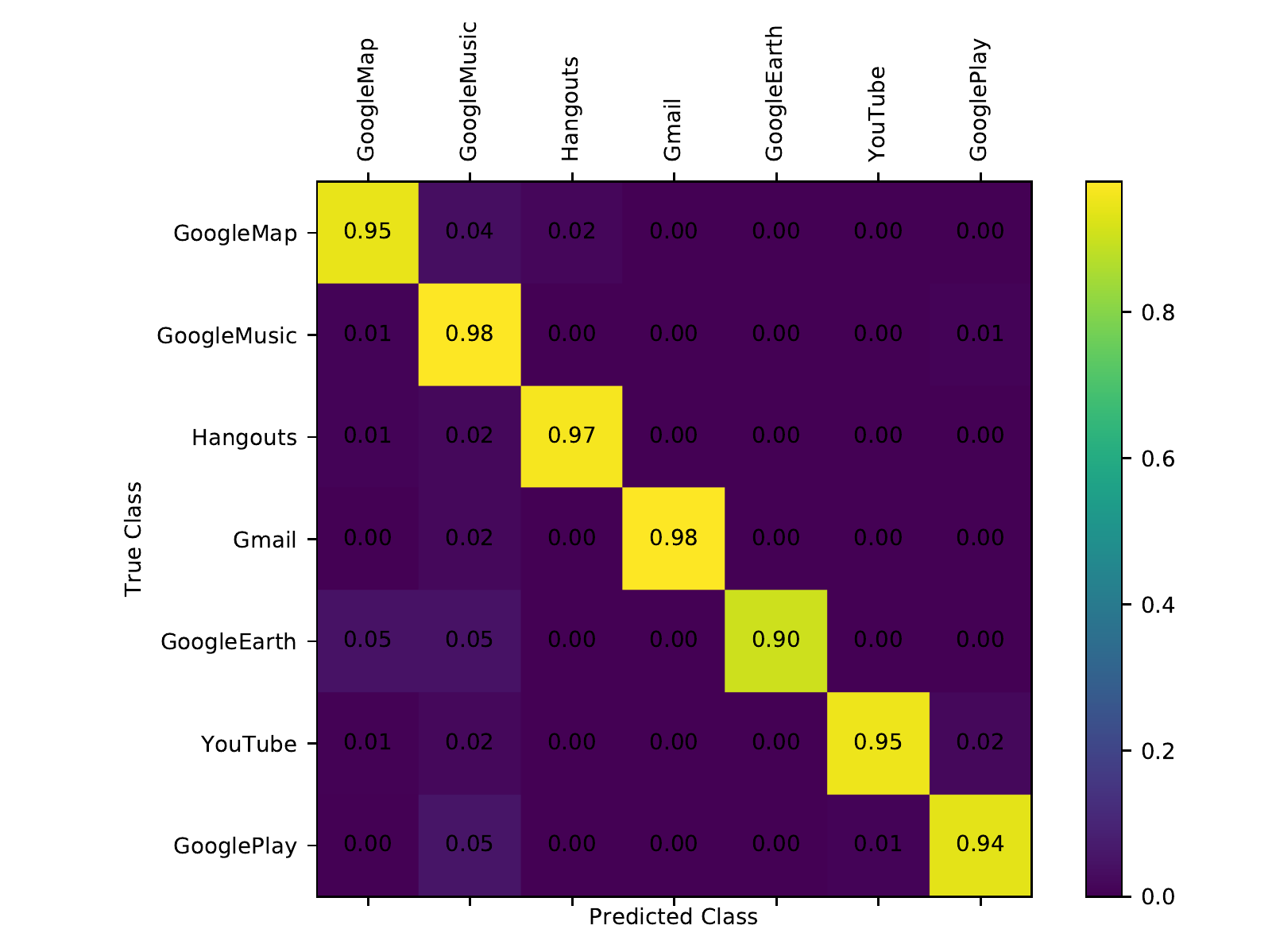}
  \caption{Confusion matrix of the multiple-flow CNN+LSTM model for flow association}
  \label{fig-cm-cnn-lstm}
\end{figure}

The confusion matrix of the 7-class app identification is illustrated in Figure \ref{fig-cm-cnn-lstm}. Google Music, Hangouts, and Gmail are the apps with the highest recall and Google earth has the lowest recall. Similar to the CNN model for single-flow classification, accuracy of HTTPS traffic is lower than unencrypted traffic. The accuracy of HTTPS traffic is $90.95\%$, as it is shown in the first row of Table \ref{tab-occlusion-lstm}.

\subsubsection{Occlusion Analysis of the CNN model for Google Services}
Table \ref{tab-occlusion-lstm} presents the occlusion analysis for Google services. Here, we only report the occluded parts that degrade the accuracy to lower than $90\%$ for brevity. The occlusion analysis of the model for 7 Google apps reveals a small difference in comparison with CNN model of 80 apps. Instead of the SNI extension, the most sensitive fields are client cipher info and extension number 18. Moreover, accuracy is not highly dependent on a single extension. But, if all extensions are occluded, the accuracy drops significantly. In both models, all extensions together are the most sensitive data field for classification of SSL/TLS traffic.


\begin{table}
  \caption{7-Google-class occlusion analysis}
  \label{tab-occlusion-lstm}
  \begin{tabular}{cl}
    \toprule
    Occluded part & Accuracy\\
    \midrule
    - & 90.95\% \\
    Record+Handshake Length & 89.30\% \\
    Rand (Server Hello) & 87.33\% \\
    Cipher info (Client Hello) & \textbf{70.64\%} \\
    All extensions & \textbf{59.14\%} \\
    Extension \#0 (SNI) & 87.31\% \\
    Extension \#16 (App layer protocol negotiation) & 88.10\% \\
    Extension \#18 (Signed certificate timestamp) & \textbf{79.23\%} \\
    Extension \#35 (Session ticket) & 89.36\% \\
    Extension \#54-65279 (Unassigned) & 89.03\% \\
    Extension \#65281 (Renegotiation info) & 85.72\% \\
    Cipher info (Client), All extensions & 32.85\% \\
  \bottomrule
\end{tabular}
\end{table}


\section{Conclusion}
\label{sec-5}
In this paper, we introduce two deep learning models for mobile app identification. We obtained a large dataset of 80 mobile apps from a large ISP in 2018. Analyzing the data revealed that around $90\%$ of traffic flows were TCP, of which around $45\%$ were encrypted with SSL or TLS. Furthermore, many apps generate not only app-specific traffic, but also ambiguous traffic that refers to common traffic, such as advertisement, that is generated by many other apps.

Using a CNN model and raw payload data of the first 6 packets of each flow without layer 2 to 4 headers, we achieve $94.22\%$ accuracy for 80-class classification for non-ambiguous traffic flows. When ambiguous traffic flows are also taken into account, the accuracy of our model is $84.03\%$ which is reasonable because identifying the source apps that generate the ambiguous traffic flows is a much more challenging task. We find that HTTPS traffic flows are the hardest one to classify. However, our model's accuracy is $75.43\%$ for HTTPS showing the feasibility of mobile app identification even for encrypted traffic in large-scale. 

We also perform occlusion analysis on SSL/TLS traffic flows to understand why these encrypted traffic flows are still classifiable. We find out that unencrypted handshake fields, including cipher info and many extensions, reveal enough information for identifying mobile apps. 

To improve the accuracy of classification, particularly for ambiguous flows, we introduce a CNN+LSTM model that takes the first 6 packets of three consecutive flows to identify the first flow. The intuition is that when an app launches, it generates several flows in a short time and some of them are ambiguous. CNN+LSTM model can look at several adjacent flows and capture the relation and order of these flows which helps improve the accuracy of traffic identification. We show that CNN+LSTM model improves the accuracy of Google app identification from $91.98\%$ with the CNN model to $96.23\%$ with the CNN+LSTM model. To the best of our knowledge, this is the first time that adjacent flows are also used to improve accuracy.

In this paper, we use payload data and a deep model to classify adjacent flows and tackle ambiguous flow problem. In the future, one plan is to study how to reduce the complexity of the model using time-series and statistical features alongside a simpler machine learning algorithm. Moreover, as more secure protocols are constantly deployed, such as TLS 1.3 which exchanges fewer unencrypted handshake fields, using a classifier that relies on unencrypted payload data may not achieve acceptable accuracy. However, using adjacent flows may help the classifier if not all adjacent flows are encrypted with a strong protocol, such as TLS 1.3. Moreover, one can study if adjacent flows using the QUIC protocol can also be identified by a similar approach. 

\nocite{*}
\bibliographystyle{IEEEtran}

\begin{IEEEbiography}[{\includegraphics[width=1in,height=1.25in,clip,keepaspectratio]{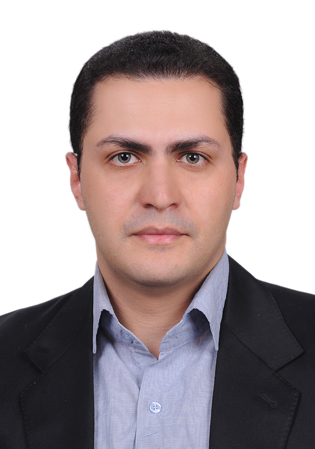}}]{Shahbaz Rezaei} (S'17)
received his B.S. degree in Computer Engineering from University of Science and Culture, Tehran, Iran, in 2011 and M.S. degree from the Sharif University of Technology, Tehran, Iran, in 2013. His research interests include machine learning security, machine learning application, mobile ad hoc networks, Mathematical Modeling and performance Evaluation as well as security and architecture of Internet. He is currently a Ph.D. student at UC Davis working on application of deep learning on computer networks and also machine learning security.
\end{IEEEbiography}

\begin{IEEEbiography}[{\includegraphics[width=1in,height=1.25in,clip,keepaspectratio]{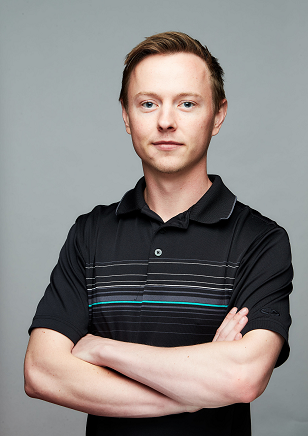}}]{Bryce Kroencke}  was born in Sacramento, California in 1997. He is currently pursuing a B.S. degree in computer science as a senior at the University of California, Davis. His involvement in professional research started in 2017 with an open-source project that involved machine learning and computational chemistry. He has completed a fellowship at Lawrence Berkeley National Laboratory, an internship at Oak Ridge National Laboratory, and is currently a student research assistant in the Engineering Department at the University of California, Davis. Bryce continues to focus his research on machine learning and its various applications.
\end{IEEEbiography}

\begin{IEEEbiography}[{\includegraphics[width=1in,height=1.25in,clip,keepaspectratio]{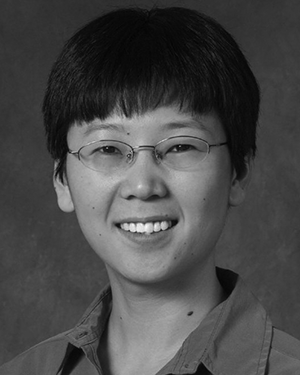}}]{Xin Liu} (M'09--F'19)
received the Ph.D. degree in electrical engineering from Purdue University, West Lafayette, IN, USA, in 2002.
She is currently a professor in the Computer Science Department, University of California, Davis, CA, USA. Before joining UC Davis, she was a postdoctoral research associate in the Coordinated Science Laboratory at UIUC. From March 2012 to July 2014, she was on leave from UC Davis and with Microsoft Research Asia. Her research interests are in the area of wireless communication networks, with a current focus on data-driven approach in networking. Dr. Liu received the Best Paper of year award of the Computer Networks Journal in 2003 for her work on opportunistic scheduling. She received the NSF CAREER award in 2005 for her research on Smart-Radio-Technology-Enabled Opportunistic Spectrum Utilization, and the Outstanding Engineering Junior Faculty Award from the College of Engineering, University of California, Davis, in 2005. She became a Chancellor's Fellow in 2011.
\end{IEEEbiography}

\EOD

\end{document}